\documentclass[11pt,twoside]{article}

\usepackage{asp2004}
\usepackage{epsf}
\usepackage[dvips]{graphicx}
\usepackage{lscape}
\usepackage{natbib}

\newcommand {\parc}[2]{\frac{\partial #1}{\partial #2}}
\newcommand {\parcsq}[3]{\frac{\partial^2 #1}{\partial #2 \partial #3}}
\renewcommand {\vec}[1]{\mathbf{#1}}

\begin{document}
\title{Introducing Powell's Direction Set Method to a Fully Automated Analysis of Eclipsing Binary Stars}
\author{A.~Pr\v sa and T.~Zwitter}
\affil{University of Ljubljana, Dept. of Physics, Jadranska 19, SI-1000 Ljubljana, Slovenia, EU}

\begin{abstract}
With recent observational advancements, substantial amounts of photometric and spectroscopic eclipsing binary data have been acquired. As part of an ongoing effort to assemble a reliable pipeline for fully automatic data analysis, we put Powell's direction set method to the test. The method does not depend on numerical derivatives, only on function evaluations, and as such it cannot diverge. Compared to differential corrections (DC) and Nelder \& Mead's downhill simplex (NMS) method, Powell's method proves to be more efficient in terms of solution determination and the required number of iterations. However, its application is still not optimal in terms of time cost. Causes for this deficiency are identified and two steps toward the solution are proposed: non-ortogonality of the parameter set should be removed and better initial directions should be determined before the minimization is initiated. Once these setbacks are worked out, Powell's method will probably replace DC and NMS as the default minimizing algorithm in PHOEBE modeling package.
\end{abstract}

\section{Introduction}

One of the most important changes in observational astronomy of the 21st Century is a rapid shift from classical object-by-object observations to extensive automatic surveys. As CCD detectors are getting better and their prices are getting lower, more and more small and medium-size observatories are refocusing their attention to detection of stellar variability through systematic sky-scanning missions. This trend is aditionally powered by the success of pioneering surveys such as ASAS \citep{pojmanski1997}, DENIS \citep{epchtein1997}, OGLE \citep{udalski1997}, TASS \citep{richmond2000}, their space counterpart {\em HIPPARCOS} \citep{perryman1997} and others. Such surveys produce massive amounts of data and it is not at all clear how these data are to be reduced and analysed. This is especially striking in the eclipsing binary (EB) field, where most frequently used tools are optimized for object-by-object analysis. A clear need for thorough, reliable and fully automated approaches to modeling and analysis of EB data is thus obvious. This task is very difficult because of limited data quality, non-uniform phase coverage and parameter interdependency. In an ongoing effort to create an automatic EB reduction pipeline, we put Powell's direction set method to the test. To the best of our knowledge, Powell's direction set method was never used before for solving the inverse problem of EBs.

\section{The Method}

Often neglected and overlooked, Powell's direction set method\footnote{Although the method is still referred to as Powell's method, the algorithm actually used is due to \cite{brent1973}, who pointed out and corrected several flaws of Powell's original algorithm.} \citep[Chapter 7]{brent1973} is a derivativeless multi-dimensional method that utilizes 1-di\-men\-si\-o\-nal minimization algorithm along a chosen direction in parameter hyperspace. It is quadratically convergent, which makes it one of the fastest methods for solving non-linear minimization problems.

The basic idea of Powell's method is to select a starting point $\vec P (p_1, \dots, p_n)$ in parameter hyperspace, and minimize the cost function along some chosen direction vector $\vec n$ using 1-D minimizer such as bracketing or Brent's parabolic method \citep[Chapter 5]{brent1973}. There is a wealth of direction set method derivatives which differ only on how and at which point of iteration is the next direction vector $\vec n$ chosen \citep{press1986}. The most appreciated scheme for choosing successive directions is the conjugate gradient method, which we now briefly describe.

When the cost function is minimized along the direction $\vec n$, the gradient in the obtained minimum is necessarily perpendicular to that direction. If it were not, that would mean that the projection of the gradient to that direction is non-null, which would in turn mean that the point was not really a minimum. The cost function may be expanded in Taylor series around that minimum:
$$
f (\vec p) = f (\vec P) + \sum_i \parc{f(\vec P)}{p_i} p_i + \frac 12 \sum_i \sum_j \parcsq{f (\vec P)}{p_i}{p_j} p_i p_j + \dots =
$$
\begin{equation}
= f (\vec P) - \vec b \cdot \vec p + \frac 12 \vec p^\dagger \cdot \vec H \cdot \vec p,
\end{equation}
where $\vec b$ is the negative gradient of $f$ and $\vec H$ is the Hessian matrix of second partial derivatives of $f$ at $\vec P$. The gradient is then simply expressed as:
\begin{equation} \label{powell_gradient}
\nabla f(\vec P) = \vec H \cdot \vec p - \vec b.
\end{equation}
From Eq.~(\ref{powell_gradient}) we may immediately deduce the change of the gradient along the chosen direction:
\begin{equation}
\delta \left( \nabla f (\vec P) \right) = \delta \left( \vec H \cdot \vec p - \vec b \right) = \vec H \cdot \left( \delta \vec p \right).
\end{equation}

To adopt the best possible direction from the found minimum $\vec P$, the method must seek an orthogonal (conjugate) direction to the former direction vector, which obviously has to point along the direction of the gradient. If the former direction is denoted with $\vec n$ and the new direction is denoted with $\vec m$, then:
\begin{equation} \label{conjugates}
\vec n \cdot \delta \left( \nabla f (\vec P) \right) = \vec n \cdot \vec H \cdot \vec m = 0.
\end{equation}

When Eq.~(\ref{conjugates}) holds for vectors $\vec n$ and $\vec m$, they are said to be \emph{conjugate}. For as long as the minimization is done \emph{only} in conjugate directions, a single minimization along a given direction is necessary, which implies quadratic con\-ver\-gen\-ce. Note that there is no need to compute the gradients at any point -- only the orthogonality condition is used. This enables Powell's direction set method to preserve its derivativeless nature.

\section{Simulation}

To assess the successfulness of Powell's direction set method for EBs, we im\-ple\-men\-ted the algorithm in PHOEBE\footnote{PHOEBE stands for PHysics Of Eclipsing BinariEs and is freely downloadable from {\tt http://phoebe.fiz.uni-lj.si}. It is based on the modeling principles of the WD code \citep{wilson1971, wilson1979}.} \citep{prsa2005}. For comparison purposes the simulation setup is identical to that of Nelder \& Mead's downhill simplex benchmarking \citep{prsa2005b}. Namely, we built a partially eclipsing synthetic main-sequence F8\,V--G1\,V binary with its principal pa\-ra\-me\-ters listed in Table \ref{params}.

\begin{table}[t]
\begin{center}
\begin{tabular}{ll}
  \includegraphics[width=5cm,height=10cm]{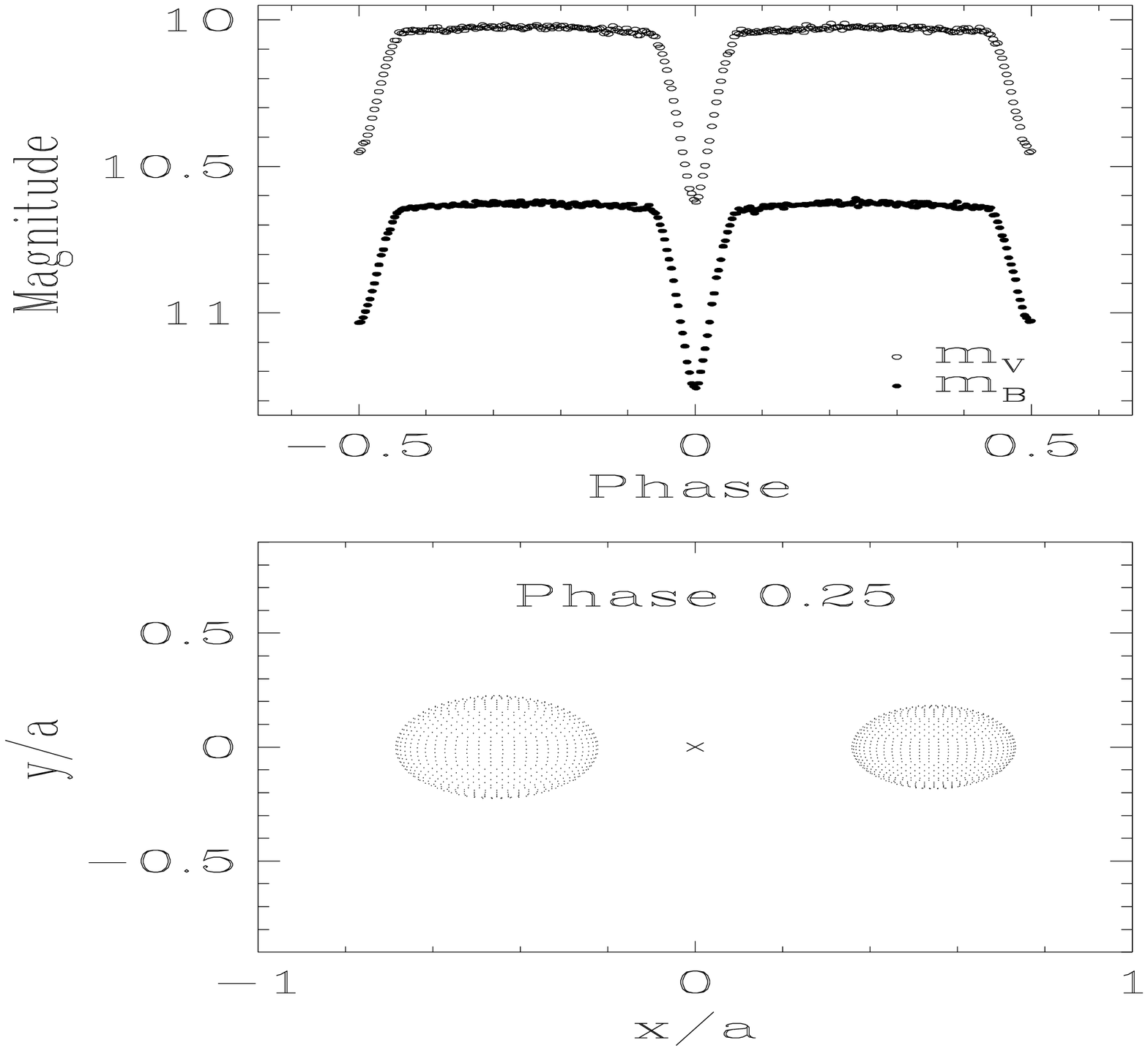} &
  \raisebox{5cm}[0pt]{
    \begin{tabular}{lrrr}
    \hline \hline
    Parameter [units] &       & Binary &       \\
                      & F8\,V &        & G1\,V \\
    \hline \\
    $P_0$ [days]                         & &  1.000 & \\
    $a\,\,\,[R_\odot]$                   & &  5.524 & \\
    $q=m_2/m_1$                          & &  0.831 & \\
    $i\,\,\,[{}^\circ]$                  & & 85.000 & \\
    $T_\mathrm{eff}\,\,\,[\mathrm K]$    &   6200 & &   5860 \\
    $L\,\,\,[L_\odot]$                   &  2.100 & &  1.100 \\
    $M\,\,\,[M_\odot]$                   &  1.236 & &  1.028 \\
    $R\,\,\,[R_\odot]$                   &  1.259 & &  1.020 \\
    $\Omega\,\,\,[-]\,{}^{\mathrm{(a)}}$ &  5.244 & &  5.599 \\
    $\log (g/g_0)\,\,\,[-]\,{}^{\mathrm{(b)}}$ &  4.33  & &  4.43  \\
    \hline \\
    \multicolumn{4}{p{6.4cm}}{${}^{\mathrm{(a)}}$~Unitless effective potentials
    defined} \\
    \multicolumn{4}{p{6.4cm}}{\ \ \ \ by \citet{wilson1979}.} \\
    \multicolumn{4}{p{6.4cm}}{${}^{\mathrm{(b)}}$~$g_0 = 1 \mathrm{cm}\,
    \mathrm{s}^{-2}$ is introduced so that} \\
    \multicolumn{4}{p{6.4cm}}{\ \ \ \ the logarithm acts on a
    dimension-} \\
    \multicolumn{4}{p{6.4cm}}{\ \ \ \ less variable.} \\
    \end{tabular}
  } \\
  \end{tabular}
\end{center}
\caption{Light curves in $B$ and $V$ passbands, star shapes and principal parameters of the simulated binary.}
\label{params}
\end{table}

In the simulation we have displaced parameters marked for adjustment (inclination $i$, temperature ratio $\tau = T_2/T_1$, potentials $\Omega_1$ and $\Omega_2$) up to $\sim 50\%$ from their correct values and started the minimization from each starting point. Attractors -- regions in parameter hyperspace that attract most convergence traces -- reveal the shape of the minimum and parameter correlations that lead to degeneracy issues. This method is known as \emph{heuristic scanning} (HS) and is described in detail by \cite{prsa2005}. During HS, only Powell's direction set method was used. Additionally, passband luminosities $L_1^i$ were computed instead of being fitted and color index constraining was used to disentangle effective temperatures of individual components (see \cite{prsa2006} for rationale and details of color index constraining).

%\section{Results}

\begin{figure}[!t]
\begin{center}
\includegraphics[width=13cm]{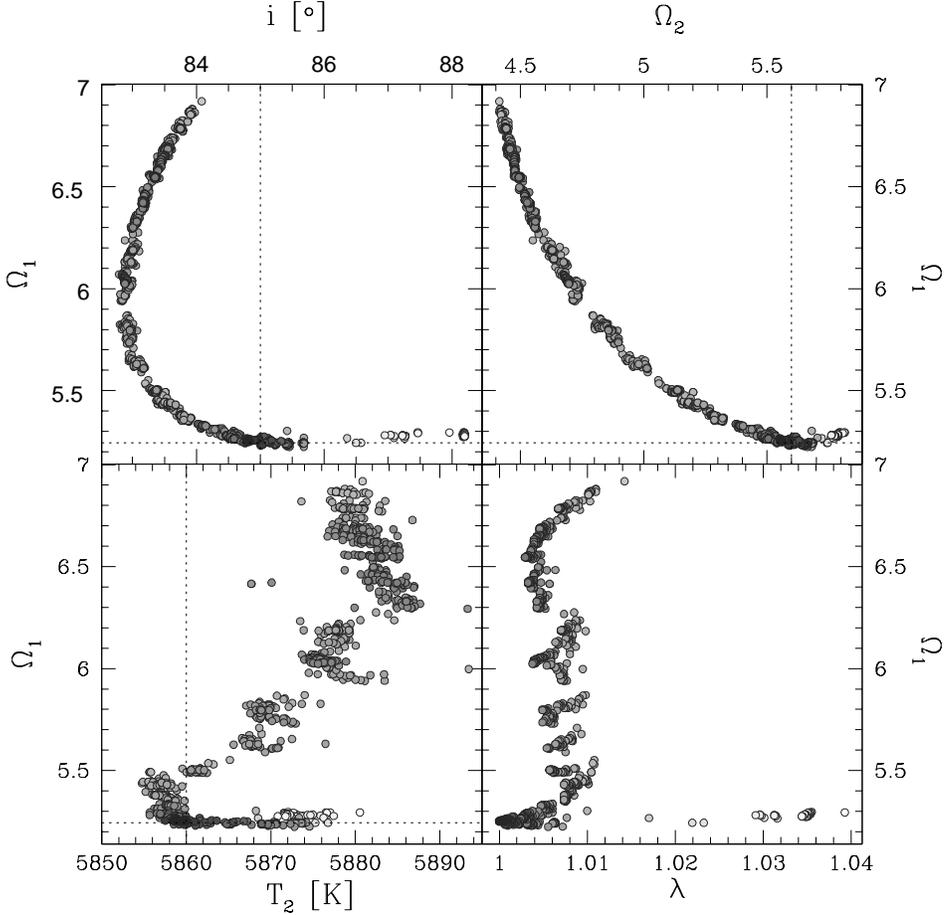} \\
\end{center}
\caption{Heuristic scan cross-sections between potentials $\Omega_1$, $\Omega_2$, inclination $i$, secondary star temperature $T_2$ and the normalized $\chi^2$ cost function value $\lambda$. Panels depict cross-sections between $\Omega_1$-$i$ (upper left), $\Omega_1$-$\Omega_2$ (upper right), $\Omega_1$-$T_2$ (lower left) and $\Omega_1$-$\lambda$ (lower right). Shades of gray symbolize the goodness-of-fit value, black being best and white being worst. Cross-hairs denote the right solution.}
\label{hs}
\end{figure}

\begin{figure}
\begin{center}
\includegraphics[width=6.5cm,height=3.25cm]{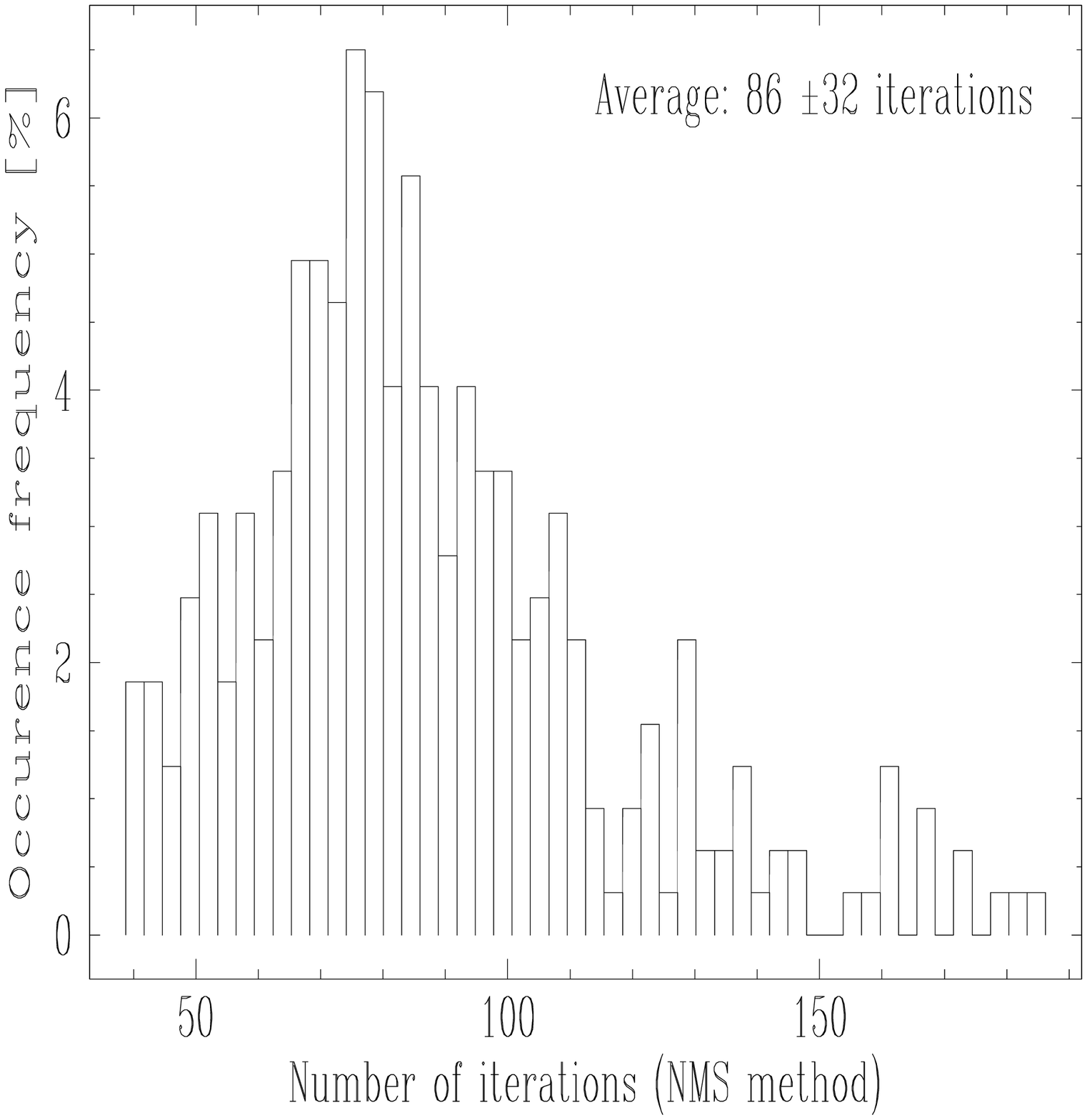}
\includegraphics[width=6.5cm,height=3.25cm]{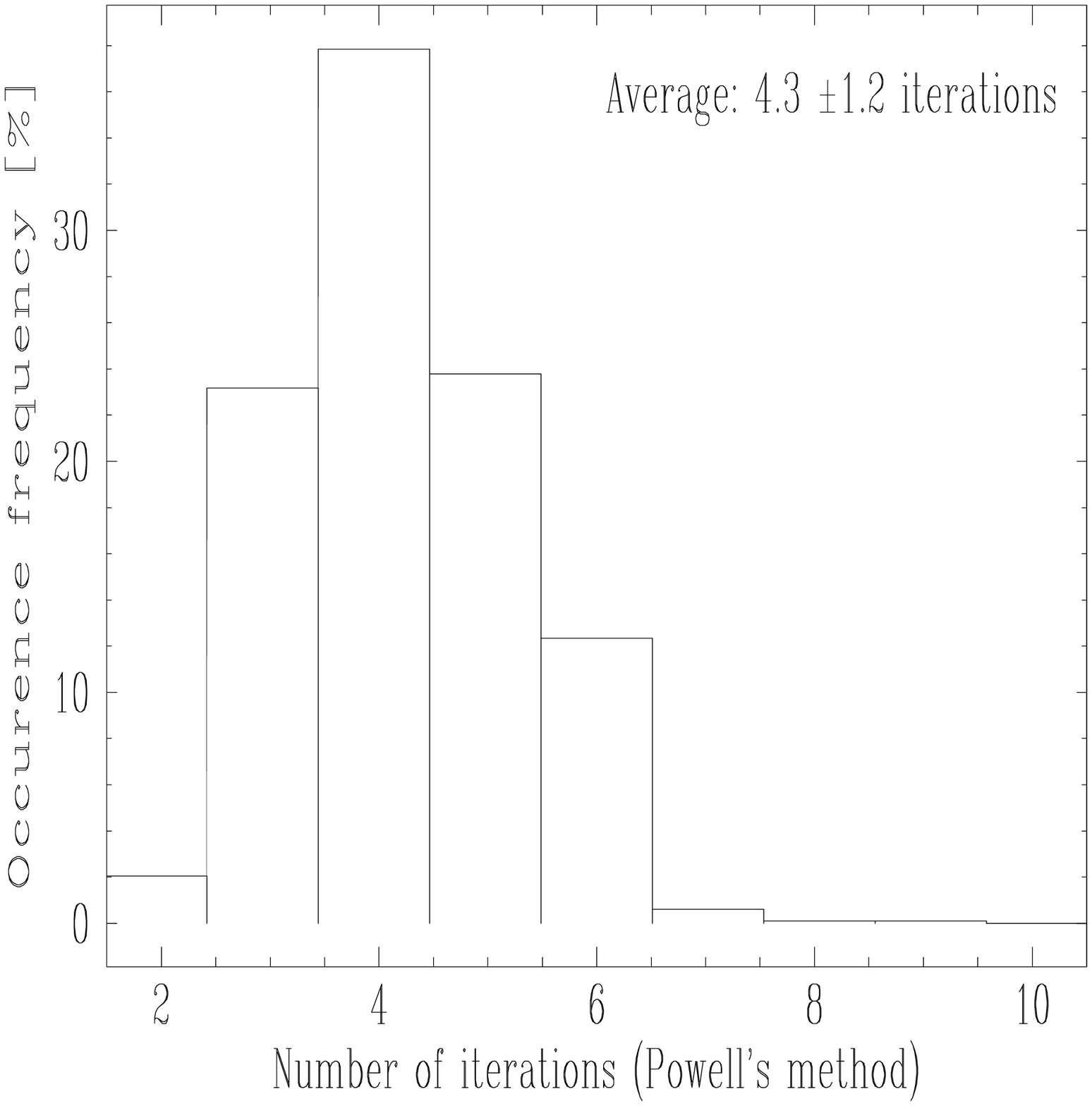} \\
\end{center}
\caption{Comparison between the number of iterations required for convergence to the fractional accuracy $10^{-3}$ in all physical parameters for Nelder and Mead's downhill simplex (NMS) method (left) and Powell's method (right). Clearly Powell's method is superior to the NMS. Note, however, that the number of iterations does not directly translate to the number of function evaluations (and thus time cost): within 1 iteration Powell's method performs line minimization along the chosen direction, which may in some circumstances imply many function evaluations.}
\label{hist}
\end{figure}

Compared to other minimization methods implemented in PHOEBE, na\-me\-ly differential corrections (DC) and Nelder and Mead's downhill simplex method (NMS), Powell's method stands out for two reasons: {\bf 1)} the solution found by HS is significantly better determined than by the other two methods. Since DC is based on computing and evaluating derivatives, it is known to diverge when parameters are not close to the minimum. NMS on the other hand always converges, but the HS solution needs to be \emph{focused} by subsequent parameter kicking (c.f.~\cite{prsa2005} and Fig.~6 therein). Powell's method is superior in this respect because the solution depicted on Fig.~\ref{hs} is determined without any need for additional focusing. Panels show the cross-sections between fitted parameters: potentials $\Omega_1$ and $\Omega_2$, the inclination $i$, secondary star's effective temperature $T_2$ and the normalized $\chi^2$ cost function value $\lambda$. Emerging shapes of the obtained minima indicate \emph{regions} (rather than individual points) that contain the solution; {\bf 2)} the required number of iterations to reach the desired fractional accuracy in parameter values is significantly lower than that of DC or NMS. Fig.~\ref{hist} depicts the comparison between the number of iterations required by the NMS method (left) and Powell's method (right). It should be stressed, however, that the number of iterations does not directly translate to the number of function evaluations: Powell's method performs line minimization in each iteration. If this minimization is performed in a direction along which the fitted parameters are correlated, it will require many function evaluations. This implies that the time cost of the overall computation does not scale with the number of iterations directly but rather depends on solution degeneracy and topology of the hyperspace. Still, given the problem at hand, Powell's method required roughly 5 times less CPU time than the NMS.

\section{Conclusion and Discussion}

Powell's method provides a solid base for further studies dedicated to solving the inverse problem for EBs in a fully automatic manner. There are two important problems that need to be worked out in order to optimize the application of this method: {\bf 1)} the non-ortogonality\footnote{By non-ortogonality we mean parameters explicitly depending on other parameters. For example, effective potential $\Omega$ explicitly depends on the mass ratio $q$ and the synchronicity parameter $F$ (c.f.~Eq.~2 of \cite{wilson1979}), yet both $q$ and $F$ appear as independent parameters in the model.} of parameter hyperspace due to classical formulation of the problem needs to be removed, which will eliminate problematic directions for line minimization and thus significantly reduce the number of function evaluations, hence the time cost; {\bf 2)} seek alternate algorithms for de\-ter\-mi\-ning subsequent sets of directions that would better suit the problem at hand. In particular, initial directions that are now set along individual parameters could be better estimated by the singular value decomposition (SVD): a method that quantifies the influence of a given parameter on the fit. This way the method could rapidly descend toward the minimum in the initial iteration already. Although Powell's direction set method seems to be most promising for automatic modeling and analysis, its time cost is still significant for large-scale surveys and needs to be furtherly reduced in future.

%\acknowledgements Support from who knows who...

\end{document}